\documentclass[letter,traditabstract]{aa} 

\usepackage{graphicx}
\usepackage{natbib}
\usepackage{txfonts}
\usepackage{epsfig}
\def\eflux{{\rm erg\,cm$^{-2}$\,s$^{-1}$}}

\def\maxi{MAXI\,J1836$-$194}
\def\inte{{\em INTEGRAL}}

\def\rxte{{\em RXTE}}
\def\swift{{\em Swift}}

\def \inte {{\em INTEGRAL}}

\def \ferg {erg cm$^{-2}$ s$^{-1}$}
\def \hcm {\hbox {\ifmmode $ atom cm$^{-2}\else atom cm$^{-2}$\fi}}

\def \apjl {ApJL}
\def \apjs {ApJS}\def \aap {A\&A}

\begin{document}
   \title{The first outburst of the black-hole candidate \maxi\  observed by \inte,\ \swift,\ and \rxte\ }

   \author{C. Ferrigno 
          \inst{1}
          \and E. Bozzo 
          \inst{1}
  \and M. Del Santo
         \inst{2}
     \and F. Capitanio
          \inst{2}        
          }

   \institute{ISDC Data Centre for Astrophysics, Chemin d'Ecogia 16,
             CH-1290 Versoix, Switzerland; \email{Carlo.Ferrigno@unige.ch}
             \and IASF-Roma, INAF, Via Fosso del Cavaliere 100, I-00152 Rome Italy
             }
   
   \date{Submitted: -; Accepted -}

\abstract{
\maxi\  is a transient black-hole candidate discovered in outburst on 30 August 2011. We report on 
the available \inte, \swift, and \rxte\  observations performed in the direction of the source during this event before 55\,864\,MJD. 
Combining the 
broad band (0.6--200\,keV) spectral and timing information obtained from these data with the results of radio observations, we show 
that the event displayed by \maxi\  is another example of  ``failed'' outburst. During the first $\sim$20 days after the onset of the 
event, the source underwent a transition from the canonical low/hard to the hard intermediate state, while reaching the 
highest X-ray flux. In the $\sim$40\,days following the peak of the outburst, the source displayed a progressive spectral hardening 
and a decrease in the X--ray flux, thus it entered again the low/hard state and began its return to quiescence.
}


   \maketitle

\section{Introduction}
\label{sec:intro}

\maxi\ was discovered in outburst by MAXI/GSC \citep{mihara11} on 30 August 2011 
(55\,803~MJD), 
and detected simultaneously also with the BAT \citep{barthelmy05} on-board 
\swift\ \citep{gehrels04}. At discovery, the flux of the source was 25~mCrab and 
40~mCrab in the energy ranges 4--10\,keV and 15--50\,keV, respectively \citep{negoro11}. 
The analysis of previous MAXI/GSC data suggested that the outburst might have already started  
on 55\,802\,MJD. Follow-up observations with \swift\,/XRT and \swift\,/UVOT on 55\,803.7\,MJD provided  
the best estimated position so far (RA=18h\,35m\,43.43s; Dec=-19d\,19m\,12.1s, J2000; associated uncertainty 
1.8~arcsec at 90\% c.l.) and led to the identification of the optical counterpart \citep{kennea11}.  
An \rxte\,/PCA observation was performed on 55\,804.5\,MJD \citep{strohmayer11}, and 
permitted a classification of the source as a new black-hole candidate (BHC) possibly undergoing a 
transition from the low-hard (LHS) to the hard intermediate state 
\citep[HIMS; see, e.g.][for a recent review]{belloni09}. 
Relatively strong radio and infrared emissions from the source were detected from 
55\,806\,MJD to 55\,827\,MJD \citep{miller11,trushkin11} and are likely associated with the presence of a jet.  

In this paper we report on all available \inte,\ \swift\,/XRT, and \rxte\,/PCA data collected 
during the outburst of the source from 55\,804 to 55\,864\,MJD.

\section{ \inte\ and \swift\ data}
\label{sec:inte}
 
\begin{figure}
\centering 
\includegraphics[width=0.9\columnwidth]{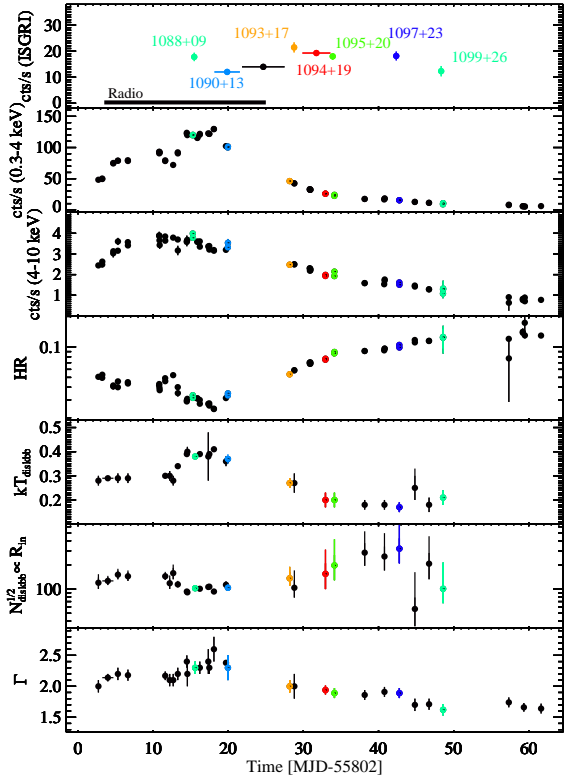}
\caption{{\it From top to bottom}.
 \inte\,/ISGRI count rate
of the source in the 20--100\,keV energy range (the Crab yields about 220\,cts/s in this band). 
Colors and labels indicate 
the time intervals used for the joint XRT+JEM-X+ISGRI fits reported in Table~\ref{tab:log}. We also 
mark in this panel the radio coverage \citep{trushkin11}.
\swift\,/XRT lightcurve of the source (time bin 1\,ks) in two energy bands 
(0.3--4\,keV and 4--10\,keV) and the corresponding HR. Evolution of the temperature and 
squared root of the normalization constant (proportional to the disk inner radius)
of the diskBB component and PL photon index.}
\label{fig:lcurve}
\end{figure}

\maxi\ was detected by IBIS/ISGRI from 
55\,816.9\,MJD to 55\,850.4\,MJD, corresponding to the satellite revolutions from 
1088 to 1099. These observations
belong to various guest observer program including the Galactic Bulge monitoring 
\citep[\texttt{http://integral.esac.esa.int/BULGE/}]{kuulkers2007}.
During this period, the source was simultaneously 
observed in the smaller field of view (FOV) of the two JEM-X units 
only for a limited amount of time (see Table~\ref{tab:log}).  
\inte\ data analysis was carried out by using version 9.0 of the OSA 
software distributed by the ISDC \citep{courvoisier03}. 
We used time bins of about three days for spectra and light curves, 
i..e., one satellite revolution, since no variability 
was detectable on shorter timescales. In Table~\ref{tab:log}, we report the  
spectral results only for those \inte\ observations in which quasi-simultaneous \swift\ data were 
available so that a broad-band fit could be carried out.
We present the IBIS/ISGRI count rate in the 20--100\,keV energy range 
in the lowest panel of Fig.~\ref{fig:lcurve}, to show the spectral
softening around 55\,820~MJD and the subsequent hardening during the outburst decay.

We analyzed all \swift\,/XRT observations performed in window timing mode (WT) 
from 55\,804.7~MJD  to 55\,863.6~MJD.
Data analysis was carried out with the technique 
described by \citet{bozzo09}. In the present case, we selected
only grade 0 events and limited our spectral 
analysis to the range 0.6--10\,keV to avoid known calibration 
problems\footnote{See http://www.swift.ac.uk/xrtdigest.shtml.}. A systematic uncertainty 
of 3\% was added to the spectra with the highest 
fluxes\footnote{See http://heasarc.gsfc.nasa.gov/docs/heasarc/caldb/swift/docs/xrt/ SWIFT-XRT-CALDB-09\_v16.pdf.}. 
\begin{table*}
\scriptsize
\centering
\caption{Quasi-simultaneous \swift\,/XRT, JEM-X, and IBIS/ISGRI spectral fits.} 
\begin{tabular}{@{}cccccccccccc@{}}
\hline
\hline
\noalign{\smallskip} 
Data${^a}$ &   \multicolumn{3}{c}{Exposures${^b}$} & $N_{\rm H} $ & $\Gamma$ & $kT_{\rm BB}$ & $10^{2}\sqrt{N_\mathrm{diskbb}}$ & \multicolumn{3}{c}{Flux${^c}$} & $\chi^2_{\rm red}$/d.o.f.\\
  &  \multicolumn{3}{c}{(ks)} & ($\times$10$^{22}$ cm$^{-2} $) & & (keV) &  $ \propto R_\mathrm{in}$ &\multicolumn{3}{c}{( $\times 10^{-9}$\,\ferg)} &  \\   
\noalign{\smallskip}
  & XRT & JEM-X & ISGRI  &  &  &  &  &   Bol & 2--10\,keV & 20--100\,keV  &   \\   
\noalign{\smallskip}
\hline
\noalign{\smallskip}
1088+09 & 1.5 & -  & 8.6 & 0.31$\pm$0.01 & 2.2$\pm$0.1 & 0.38$\pm$0.01 & $1.0\pm0.3$ &  12.1& 1.6  & 0.7 & 1.13/397\\ 
\noalign{\smallskip}
1090+13 & 0.7 & 18.0  & 56.5 & 0.33$\pm$0.01 & 2.24$\pm$0.03 & 0.37$\pm$0.01 & 1.0$^{+0.4}_{-0.1}$ & 10.6 & 1.3 & 0.6 & 0.99/329 \\ 
\noalign{\smallskip}
1093+17 & 0.7 & - & 8.2 & 0.24$\pm$0.01 & 1.95$\pm$0.04 & 0.28$\pm$0.02 & 1.1$^{+0.8}_{-0.6}$ & 6.6 & 0.8 & 0.9 & 0.98/239 \\ 
\noalign{\smallskip}
1094+19 & 1.0 & 12.0 & 67.7 & 0.23$\pm$0.02 & 1.91$\pm$0.02 & 0.21$\pm$0.02 & 1.3$^{+1.5}_{-1.0}$ & 4.8 & 0.6 & 0.8 & 0.98/287\\
\noalign{\smallskip}
1095+20 & 0.6  & - & 13.2 & 0.27$\pm$0.01 & 1.84$\pm$0.05 & 0.21$\pm$0.04 & 1.4$^{+2.5}_{-1.1}$ & 5.7& 0.7 & 1.0 & 0.92/202 \\
\noalign{\smallskip}
1097+23 & 1.0  & - & 9.9 & 0.33$^{+0.04}_{-0.10}$ & 1.85$\pm$0.06 & 0.18$\pm$0.03 & 2$^{+4}_{-2}$  & 4.2 & 0.5 & 0.7 & 0.94/205\\
\noalign{\smallskip}
1099+26 & 1.0  & - & 10.2 & 0.26$\pm$0.07 & 1.5$\pm$0.1 ($E_{c} > 53$ keV) $^d$ & 0.25$\pm$0.06 & 0.6$^{+1.0}_{-0.5}$ & 2.9 & 0.4 & 0.9 & 0.90/166\\
\noalign{\smallskip}
\hline
\noalign{\smallskip}
\multicolumn{12}{l}{$^a$: Indicates the \inte\ revolution + the latest two digits XX of the \swift\ observation 000320870XX.}\\  
\multicolumn{12}{l}{ $^b$: effective exposure time of XRT, JEM-X (both units), and ISGRI, respectively. $^c$: unabsorbed fluxes. ``Bol'' indicates the model flux in the 0.1--200\,keV energy range.}\\
\multicolumn{12}{l}{$^d$ This spectrum is modeled by a cut-off power-law with $E_c=150$\,keV.}
\end{tabular}
\label{tab:log}
\end{table*} 
The XRT light curves of the source in the 0.3--4\,keV and 4--10\,keV energy bands, together with the corresponding hardness ratio (HR),
are reported in Fig.~\ref{fig:lcurve}.   
None of the XRT spectra extracted before 55\,850\,MJD could be satisfactorily fit 
by using an absorbed PL component. The residuals from these fits demonstrated 
the presence of an additional soft spectral component below $\sim$4\,keV. 
A better fit to the data was obtained by adding a 
disk blackbody component (\textsl{diskBB} in {\sc Xspec}). 
The XRT spectra extracted after 55\,850~MJD did not require the soft component and could be well 
fitted by using an absorbed PL (see Fig.~\ref{fig:lcurve}). 

\begin{figure}
\label{fig:broadband} 
\begin{center} 
\includegraphics[width=0.75\columnwidth,angle=90]{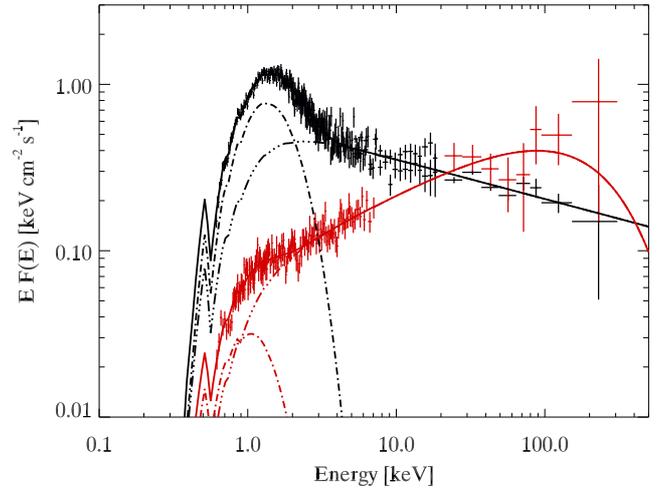}
\caption{Broad-band spectrum of \maxi\ obtained by using data from 
XRT, ISGRI, JEM-X1 and JEM-X2 (the model comprises PL and diskBB 
components). We show in black the data 
1090+13 (see Table~\ref{tab:log}) of the outburst peak.
Red points are from the observations 1099+26 at the end of the outburst 
and are modeled introducing
an exponential roll-over with $E_c=150$\,keV.} 
\end{center}
\end{figure}

We performed a joint fit of all available quasi-simultaneous XRT+ISGRI data 
to investigate the broad-band (0.6--200\,keV) spectral properties 
of the source (see Fig.~\ref{fig:broadband}). The results of this analysis are reported in Table~\ref{tab:log} (throughout the paper uncertainties are estimated at 90\% c.l.). 
When available, we included in the fit the spectra extracted from the two JEM-X units. 
A normalization constant on the order of unity 
was introduced to account for the variability of the source 
and inter-calibration issues between the different instruments. 
Every broad-band spectrum 
could be reasonably well fitted by using an absorbed diskBB+PL model. 
No high-energy exponential roll-over was observable:  
a lower limit on the cut-off energy could be generally set at $\sim$200\,keV, with the exception of 
the spectrum 1099+26, for which the relatively poor statistics of the data 
yields a less constraining lower limit of 53\,keV. 

\section{ \rxte\ data}
\label{sec:rxte}

\rxte\,/PCA observations (55\,804.5--55\,864\,MJD)
were analyzed with standard tecniques, as described in \citet{ferrigno11}. 
We used only data from the PCU2 because this was always active throughout 
the outburst of \maxi\ (observations ID~96438 and 96371; total exposure time 62\,ks). 

\begin{figure}
\label{fig:spectra}
\includegraphics[width=0.9\columnwidth]{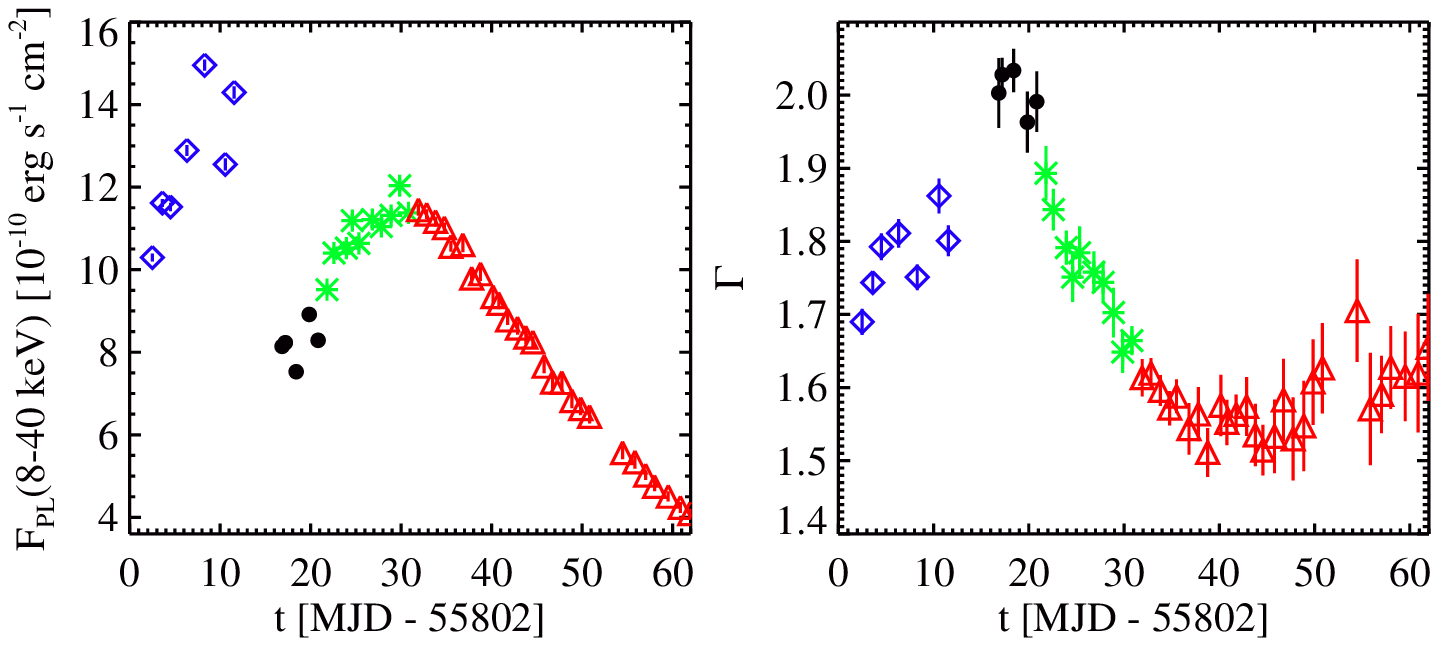}\\
\includegraphics[width=0.9\columnwidth]{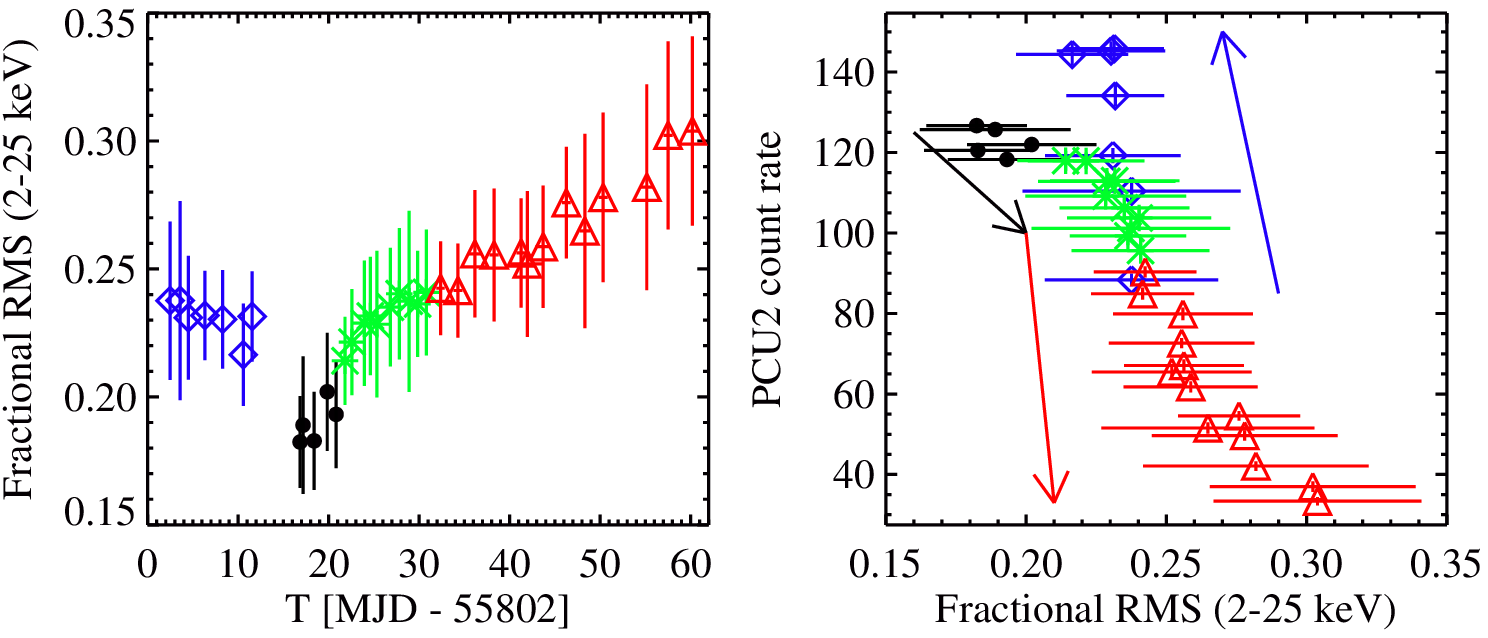}
\caption{{\it Upper panels}: spectral model parameters obtained from the PCA data in the 8--40\,keV energy range; $F_\mathrm{PL}(8-40\,\mathrm{keV})$ is the 8--40\,keV flux 
in units of $10^{-10}\,\mathrm{erg\,s^{-1}\,cm^{-2}}$, $\Gamma$ the PL photon index. 
Different symbols mark in all panels the three phases of the oubursts 
discussed in Sect.~\ref{sec:rxte}. {\it Bottom left panel}: fractional RMS computed from the background subtracted RXTE/PCA light curves in the 2--25\,keV energy range as a function of time. \emph{Bottom right panel}: fractional RMS vs. intensity. 
The average count rate is background-subtracted and refers to PCU2. 
Arrows indicate temporal evolution.}
\end{figure}

The spectral analysis was performed using the \texttt{standard2} mode 
data in the energy range 8--40\,keV, 
as the contamination from the Galactic ridge prevented an accurate 
characterization of the source spectral emission properties at lower energies (we relied on XRT data for 
the energy range 0.6--8\,keV).
In the considered energy range, the PCA spectrum could be modeled by a power-law with 
$\Gamma$$\sim$1.5--2.0 (see Fig.~\ref{fig:spectra}, upper panels). The PL 
photon index increased ($\Gamma$$\simeq$1.7-1.9) 
during the earlier phases of the outburst (55\,803--55\,817\,MJD, blue diamonds), 
became steeper ($\Gamma$$\simeq$2.0) 
around the peak of the outburst (55\,817--55\,822\,MJD, black dots), decreased until 
55\,838\,MJD, and remained roughly constant afterwards. 
The flux of this PL
component was suppressed at the outburst peak (black dots), 
increased during the spectral hardening until $\sim$55\,832\,MJD (green stars), and finally decreases steadily (red triangles). 
These results agree fairly well with those obtained from \swift\ and \inte.\  

\begin{figure}
\includegraphics[width=0.48\columnwidth]{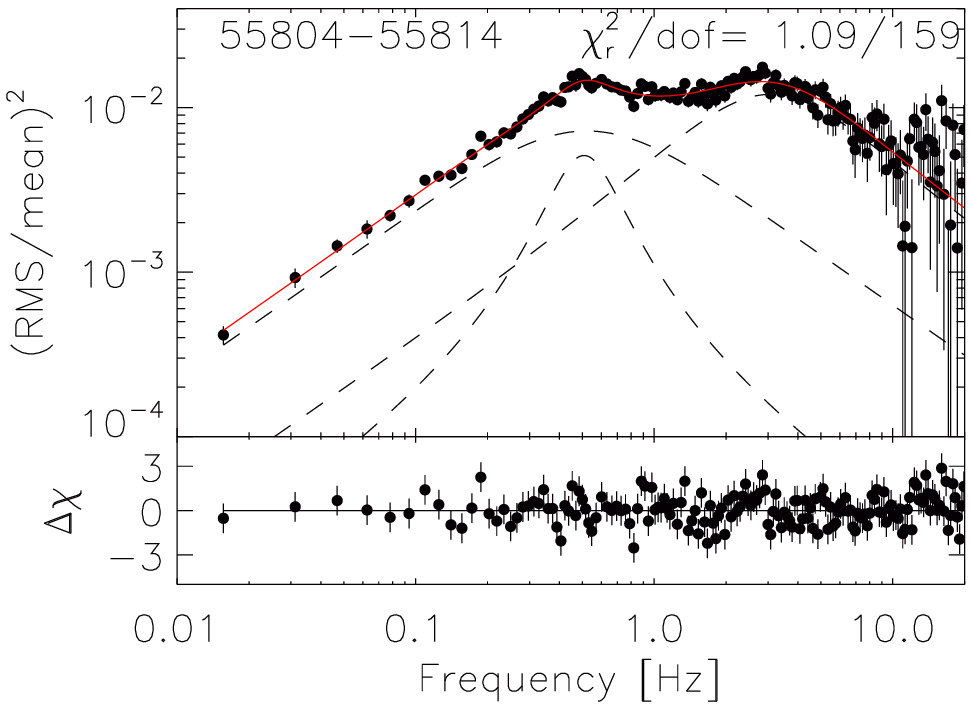}\hfill
\includegraphics[width=0.48\columnwidth]{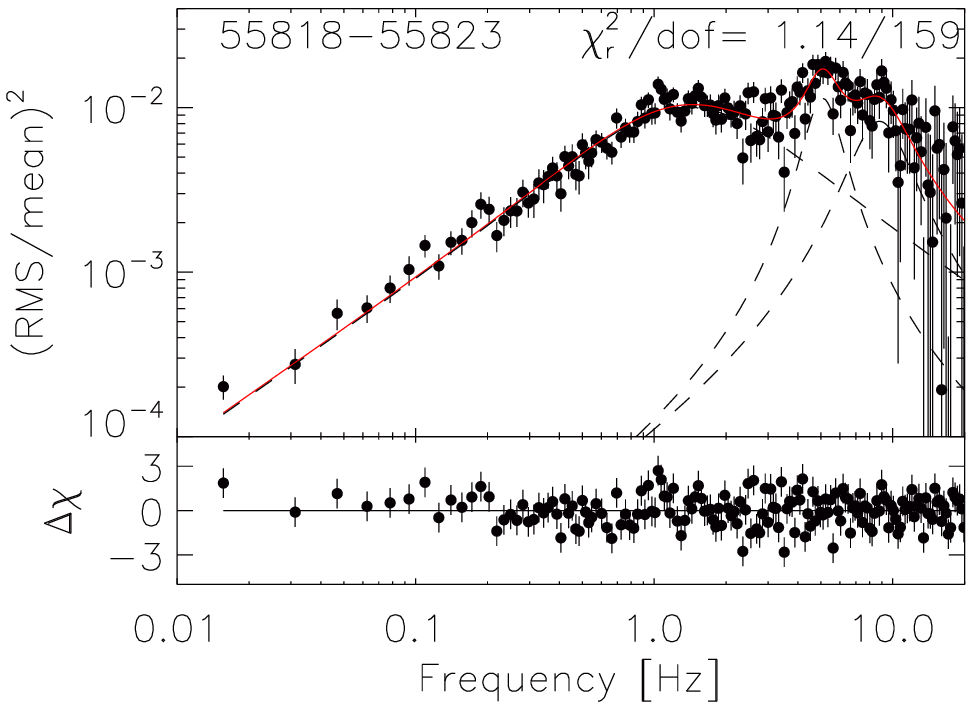}\\
\includegraphics[width=0.48\columnwidth]{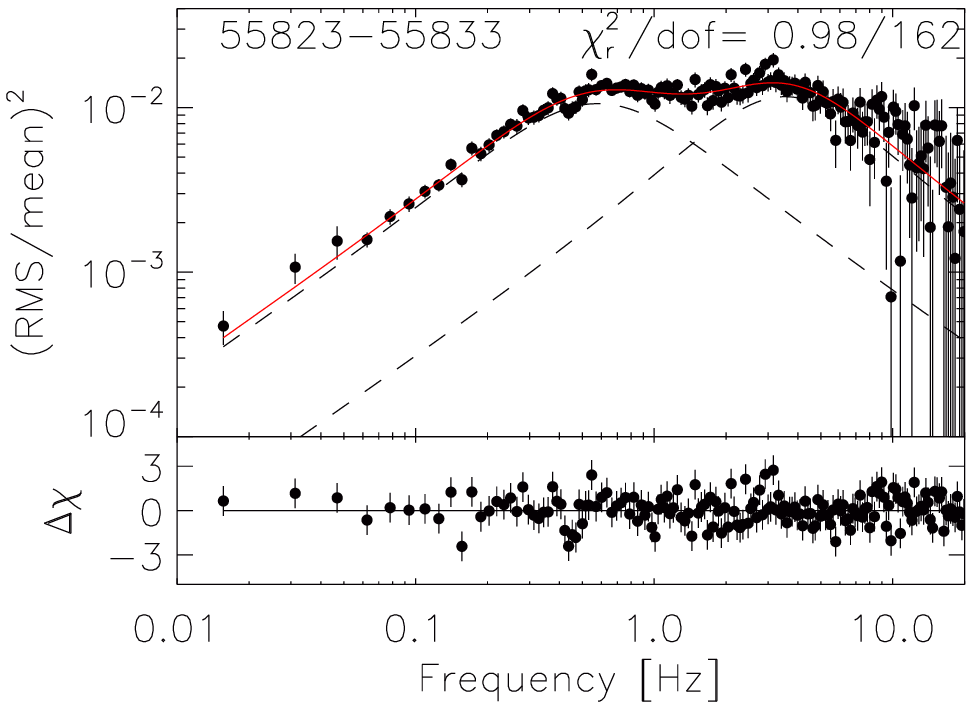}\hfill
\includegraphics[width=0.48\columnwidth]{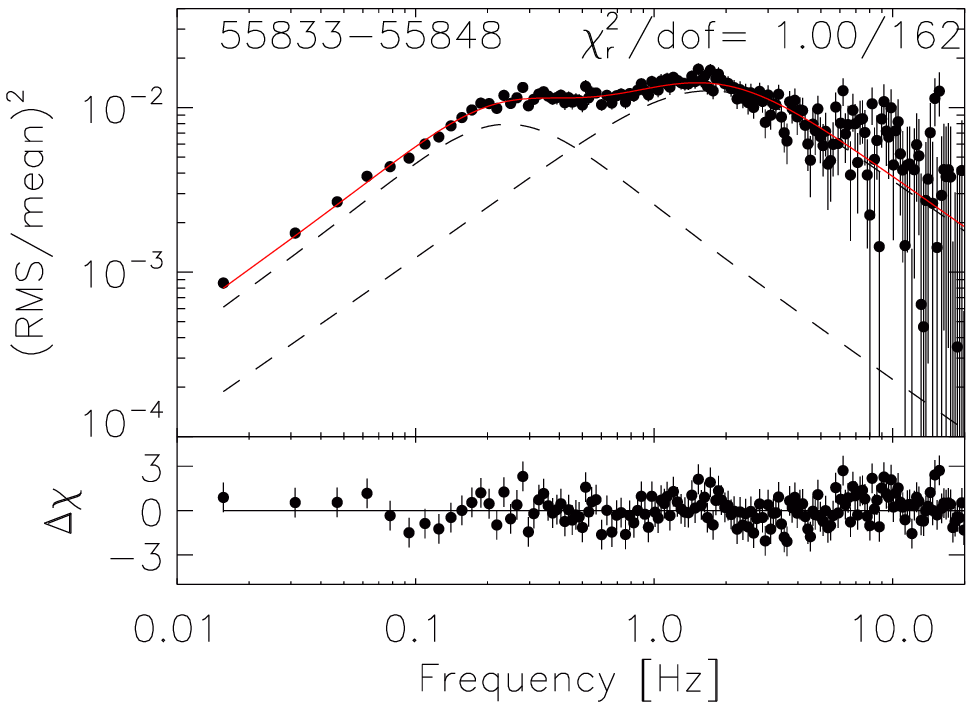}
\caption{Four PSDs extracted from the RXTE/PCA observations of \maxi\ (see Sect.~\ref{sec:rxte}). The red solid line is the best-fit model obtained from the sum of
the Lorentzian curves plotted as dashed lines. Residuals form the fit are shown below each PSD.}
\label{fig:psd}
\end{figure}

The timing analysis was performed using the \texttt{GoodXenon} events in the energy range 2--25\,keV.
Power spectra (PSD) were calculated using \texttt{powspec v1.0} during intervals of 64\,s,  
averaged together, and geometrically rebinned.
In preliminary runs, we used $2^{-10}$\,s time bins and verified that above $\sim$30\,Hz the signal 
was always consistent with white noise. The latter, also corrected for the dead-time contribution, 
was subtracted from the final PSD, which we
computed from the background-corrected light curves binned at $2^{-6}$\,s.
We averaged the PSDs in long intervals corresponding to the different phases of the outburst  
and verified that the PSD of the 
single observations within these intervals presented similar characteristics.     
As shown in Fig.~\ref{fig:psd}, all PSDs are characterized by a flat-top, band-limited noise that can be described using Lorentzian curves (Fig.~\ref{fig:psd}). 
The PSD is dominated by two broad components whose centroid frequencies move coherently throughout the outburst in rough correlation with the source flux. 
In the early phase of the outburst (55\,804-55\,814~MJD) the lower component peaks at $\sim$0.5\,Hz,
while at the maximum of the outburst (55\,818-55\,822\,MJD) 
it moves to $\sim$1\,Hz.
We note that the higher frequency component splits at the outburst peak into narrower features centered at $5.0\pm0.1$\,Hz and $8.4\pm0.6$\,Hz, with quality factors of $5\pm1$ and $3\pm1$, respectively (fractional RMS of $\sim8$\%, uncertainties at $1\sigma$). Integrating the PSD on shorter intervals, we were unable to verify wether these features are produced by narrower peaks moving in frequency, owing to the limited statistics.
At the later phases, we averaged the PSD over the intervals 55\,823-55\,833~MJD and 55\,833-55\,848~MJD to study possible variations  of the PSD,
to show a progressive decrease of the noise peak frequency (we neglected the following observations because of the lower S/N).

In  the lower panels of Fig.~\ref{fig:spectra},
we show the integrated fractional RMS computed
from the background-subtracted light curves 
using bins of $2^{-4}$\,s as $\left(\sum_{i=0}^{N}(r_i-\bar r)^2 - \sum_{i=0}^{N} \sigma_i^2\right)^{1/2} / (N \bar r)$ ($r_i$ is the rate in the $i$-th bin with error 
$\sigma_i$, $N=2048$ is the number of bins, and $\bar r$ is the mean count rate). 
The RMS of each observation and the corresponding uncertainty was estimated from  
the average and standard deviation of the values determined in each 64\,s long time intervals. The fractional RMS remains above 21\% at the 
early and late stages of the outburst, while it is slightly suppressed ($\sim$18\%) at the peak (black dots)\footnote{We checked that similar results would have been obtained from 
the timing analysis of the XRT observations, though with larger uncertainties 
owing to the limited statistics of the data.}. 

\section{Discussion}
\label{sec:discussion}
\begin{figure}
\centering 
\includegraphics[width=0.9\columnwidth]{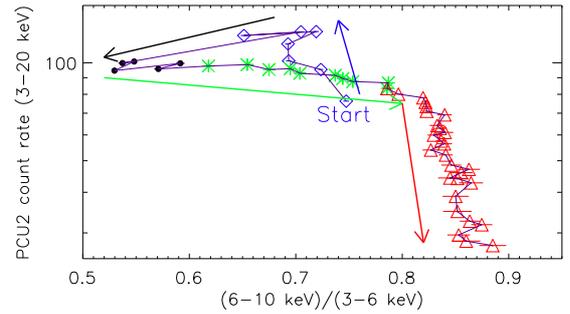}
\caption{
Hardness intensity diagram (HID) of \maxi\  obtained from the \rxte/PCA background-subtracted 
average spectra of each observation. Symbols refer to the same intervals as in Fig.~\ref{fig:spectra}, arrows indicate temporal evolution.}
\label{fig:hid} 
\end{figure}

When undergoing an outburst, BHC sources usually follow a relatively well 
known ``q-track'' pattern in the HID \citep[see Fig.~\ref{fig:hid};][]{Fender04,Homan05,belloni09}. 
The outburst starts in the so-called low-hard spectral state (LHS), which is characterized by 
a power-law shaped X-ray spectrum with $\Gamma$$\simeq$1.4-1.5 and a cut-off at the higher energies 
($E_{\rm cut}$$\sim$100\,keV). A soft thermal component with $kT<0.5$\,keV is often observed in the LHS 
and ascribed to the emission from an accretion disk truncated at long distances from the central BH 
\citep[$\sim$100~km;][]{done07}. The radio emission in this state is caused by the synchrotron radiation 
of a steady jet. During the early rise of the
outburst, the X-ray and radio luminosities both increase, but the
X--ray color of the spectrum remains hard \citep{Corbel00,Corbel03}.  
As the outburst progresses, the source reaches the high-soft state (HSS) characterized by a prominent 
thermal emission from the accretion disk and a marginal power law tail.
In addition to these two main states, \citet{Homan05} identify the hard-intermediate (HIMS) and soft-intermediate (SIMS) states with spectral parameters of the disk and power-law component in between the LHS and HSS. We will exploit this classification in the remainder of this discussion. An alternative division of the source intermediate states was used, e.g., by \citet{Remillard06}, who define a steep power-law state ($\Gamma>2.4$) generally observed at high flux and then also known as very high state, plus an intermediate state that covers the unclassified observations.
We stress that both SIMS and HSS are characterized by the absence of radio emission. This is interpreted as the suppression of the jet and marks the most striking difference between the soft and hard states.
The short timescale variability 
also varies during the outbursts.  The level of the
RMS noise decreases with increasing flux in the LHS and decreases even more
when leaving this state. 
Band-limited noise is then suppressed and
quasi-periodical oscillations (QPOs) make their appearance \citep[see e.g.,][]{Homan05,vanderKlis06,Remillard06, belloni09}. 
Not all transient BHCs in outburst go through a complete q-track.
So far, a limited number of objects were observed 
to start an outburst, reach the HIMS, but then 
return to the LHS instead of moving to HSS \citep{fiamma09}

The available data of the 2011 outburst of \maxi\ suggest that this event  
represents another example of these ``failed'' outbursts.  
As summarized in Sect.~\ref{sec:intro}, the initial observations of the source 
carried out with MAXI indicate that the onset of the outburst occurred probably 
on 55\,802~MJD. Spectral and timing information on the source were first
available through \swift/XRT and RXTE/PCA 
observations in the interval 55\,803--55\,814~MJD, showing characteristics 
typical of the end of LHS: 
a relatively hard power-law photon index increasing from $\sim$1.7 to 2.0, 
a relatively high RMS ($\sim$23\%, Fig.~\ref{fig:spectra}), and a marginally significant QPO corresponding to the break frequency of the PSD. 
A soft diskBB component was detected in the XRT spectrum, with a temperature $\sim$0.3\,keV, i..e,
comparable with that expected from a truncated accretion disk in LHS \citep{done07}. 

During the following bright phase of the outburst (55\,815--55\,822~MJD), 
the diskBB accounts for about half the broad-band X-ray flux (Fig.~\ref{fig:broadband}) with an increased temperature of $\sim0.4$\,keV and a roughly halved squared root of normalization, which indicates a smaller inner truncation radius of the disk (Fig.~\ref{fig:lcurve} and Table~\ref{tab:log}). 
Correspondingly, the power-law steepens ($\Gamma\sim$2)
consistently with a more efficient cooling of a 
population of high-energy electrons, which
up-scatter the soft disk photons and produce the high-energy non-thermal emission
\citep[see, e.g.,][and references therein]{zdziarski2002}.
These properties indicate a transition to the HIMS, which is confirmed by a decrease of the
fractional RMS to $\sim$18\%,
the higher frequency extension of the broad-band noise, and the appearance of
blandly coherent noise at a few Hertz. 
The latter often evolves into QPOs with high quality factors 
during the HIMS \citep[see, e.g.,][]{Homan05}; this is not observed for \maxi.
A transition to the SIMS is excluded by the absence of a clear drop in the RMS to a few percent (see Fig.~\ref{fig:spectra}) and by radio detection of the source 
from $\sim$55\,806 to $\sim$55\,827\,MJD \citep[this argues against any possible disappearence of the jet, as expected in the transition HIMS-SIMS;][]{trushkin11}.   

After 55\,823~MJD, \maxi\ underwent a relatively rapid flux decrease, 
a significant spectral hardening and an increase of the fractional RMS (see Fig.~\ref{fig:spectra}). 
The hard X-ray flux as measured by RXTE/PCA
resumed during the first part of the decay (until $\sim$55\,832~MJD) and then
decayed at roughly constant spectral slope.
Correspondingly, the temperature of the thermal component decreased to $\sim$0.2\,keV (Fig.~\ref{fig:lcurve}) and its normalization increased: this can be interpreted, within the disk truncated model, as the inner radius moving away from the BH.
\swift\ data collected in the period from 48 to 61 days after the onset of the event did not show evidence for the soft spectral 
component probably because of the limited and short exposure time of the \swift/XRT data and/or a very low disk temperature. In this phase, the PL photon index remained virtually constant at $\Gamma$$\simeq$1.6, and the source became fainter down to an X--ray flux of 3$\times$10$^{-10}$\,\eflux. 
 
The HID of \maxi\ showed in Fig.~\ref{fig:hid} also supports these conclusions: the source was first observed during the transition from the LHS to HIMS (blue diamonds). It remained in the HIMS for a few days (black dots). It then slowly moved back to the LHS (green stars) and then faded out 
(red triangles).
Based on the timing and spectral properties of  the \rxte\ , \swift\ , and \inte\ data, we conclude 
that \maxi\ is one of the few black-hole candidates, together with  H\,1743$-$322 in 2008 and SAX\,J1711.6$-$3808 in 2001 \citep[see][and references therein]{fiamma09},
which showed a transition to the HIMS, but did not enter a soft state.
\begin{acknowledgements} 
M.D.S. and F.C. acknowledge financial contribution
from the agreement ASI-INAF I/009/10/0. M.D.S. acknowledges
the grant from PRIN-INAF 2009 (PI: L. Sidoli). We thank P. Casella for 
sharing his experience in interesting discussions.
\end{acknowledgements}

\bibliographystyle{aa}
\bibliography{maxi1836}

\end{document}